\documentclass[runningheads]{svmult}
\usepackage{makeidx}   
\usepackage{graphicx}  
\usepackage{subeqnar}  
\usepackage{multicol}  
\usepackage{cropmark} 
\usepackage{physprbb}  
\makeindex             
\newcommand{\greeksym}[1]{{\usefont{U}{psy}{m}{n}#1}}

\begin{document}
\title*{Preon Prophecies by the Standard Model}
\toctitle{Preon Prophecies by the Standard Model}

\titlerunning{Preon Prophecies}

\author{Sverker Fredriksson}
\authorrunning{Sverker Fredriksson}
\institute{Department of Physics, Lule\aa\ University
of Technology, SE-97187 Lule\aa, Sweden}

\maketitle

\begin{abstract}
The Standard Model of quarks and leptons is, at first sight, nothing but
a set of {\it ad hoc} rules, with no connections, and no clues to
their true background. At a closer look, however, there are
many inherent prophecies that point in the same direction:
{\it Compositeness} in terms of three stable preons.
\end{abstract}

\section{Introduction}
The Standard Model (SM) of quarks and leptons serves as a bible
of high-energy physics. It even resembles the
real Bible, in the sense that it is mainly a set of ``stories'',
rules and wisdoms, which reflect both real life and human thinking.

The SM is also flexible, since it can easily be complemented with
new findings or ideas, such as neutrino oscillations, supersymmetry,
or new fundamental quarks and leptons, just like the first Bible
was once complemented with the New Testament by Christians.
Unlike the Bible, the SM does not give a clue to
its deeper meaning, nor does it reveal any connections between
its many bits and pieces. The SM is hence a quantitative success,
but a qualitative mess.

What I will argue in this talk, however, is that the situation is not
that unclear or hopeless. Rather, the SM is full of prophecies and hints
to its deeper background, many of which can be understood after
a deeper look into those SM features that have been put in entirely
by hand. I will make frequent references to the
``History Book'', in order to investigate if such ideas or situations have
appeared before. As you will see, there have been numerous situations in the
past when paradoxes and problems like the ones in the SM turned out
to come from {\it compositeness} of the ``fundamental'' particles
of those days.

To be more precise, I will discuss the following
aspects of the SM:
\vskip0.1cm
- there are six quarks and six leptons in three families

- most quarks and leptons are unstable

- there are several {\it ad hoc} quantum numbers

- some quarks, leptons and gauge bosons mix/oscillate

- the heavy vector bosons are massive and unstable

- the $Z^{0}$ mixes with the photon
\vskip0.1cm
The discussion will converge toward more
precise features of a preon model, and I will conclude by
briefly describing a recent one, presented by Jean-Jacques
Dugne, Johan Hansson and myself \cite{dugne02}. The discussion
presented here is, in fact, an offspring of our work with
that particular model.

\section{Some General Observations}
Before going into some detail with preons and the SM,
a few more general observations can be made. First of
all, the History Book tells us that ``there is
always a deeper layer of compositeness''. Earlier layers have
normally been suggested and/or discovered when the model
in fashion became too complex, or when there were too many
different models for the ``fundamental'' particles.
Such leaps to new levels of compositeness come
every $30-40$ years, which means that time
is ripe for preons!

Nevertheless, compositeness has never been a main
trend of high-energy research. Rather, an
overwhelming majority of theorists has traditionally
regarded the elementary particles of the day as
{\it the} fundamental ones.

Compositeness usually comes into fashion as a viable
theory only if strongly supported by experiment,
or if suggested by some well-known theorist. In the latter
case, the interest fades away after some time, if
not supported by observations.
Examples are the quark model by Gell-Mann and Zweig
\cite{gellmann64,zweig64} and the early preon
models by Harari \cite{harari79} and Shupe \cite{shupe79}
(``rishons'') and by Fritzsch and
Mandelbaum \cite{fritzsch81} (``haplons'').

The quark model gained a wider acceptance only
after the pioneering deep-inelastic scattering
data of the late 1960s and the early 1970s, and
their interpretation in terms of partons. There
are no such supporting data for preons, and the
interest in the first preon models was therefore
gone a long time ago. The concept of compositeness still
appears now and then in experimental work, but mostly
in routine searches for deviations from the SM
predictions in events with large transferred momenta.
The lack of signals are normally quoted as a minimal
preon energy scale of a few TeV \cite{pdg02}. It is to be
understood as an inverse length-scale in a hypothetical
form factor of internal preon wave functions (and
not as a preon mass scale). In order to restrict
this length-scale further, one would seemingly
need the high energies of the upcoming CERN LHC
facility, or at least high-statistics data
from RUN II of the Fermilab Tevatron.
However, this does not exclude that preons
could reveal themselves in other ways, e.g.,
at the Tevatron, such as through the discovery of
new, exotic quarks or leptons. An earlier example
was the discovery of the exotic $\Omega^{-}$ hyperon
and its importance for the subsequent quark model.

\section{Too Many Quarks and Leptons}
The most common argument in favour of preons \cite{souza92}
is that there are too many quarks and leptons to
let any SM enthusiast feel comfortable. There is
no obvious logical reason why there would be (at least)
twelve fundamental particles. The case for preons is
strengthened by the fact that these particles
fall into a nice pattern of three ``families''.

\begin{figure}
\begin{center}
\includegraphics[width=.7\textwidth]{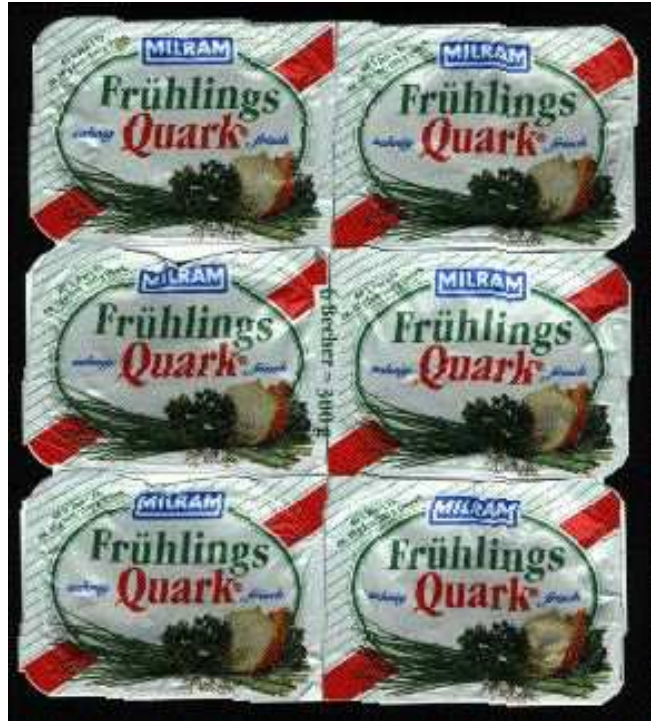}
\end{center}
\caption[]{Why do quarks come in six-pack, like in
German supermarkets?}
\label{eps1}
\end{figure}

The History Book tells us that the existence of
too many fundamental particles in the past always
preceded the discovery of yet smaller and fewer building blocks
(not counting the era of ``earth, fire, air and water'').

Examples are:
\vskip0.1cm
- the existence of too many elements was explained by the atoms
consisting of electrons and nuclei;

- the existence of many isotopes of one and the same element
is a by-product of the compositeness of nuclei, containing
protons and neutrons;

- the hundreds of hadrons that puzzled high-energy physicists
a half-century ago reflect their quark structure.
The patterns among the early hadrons led the mind to
the number three, and resulted in a model with three
different quarks, with a quark-flavour $SU(3)$ symmetry.
\vskip0.1cm
A first conclusion is therefore that the many quarks
and leptons reflect a preon substructure, and
that the pattern of three families comes from the
existence of three preons, with a preon-flavour
$SU(3)$ symmetry. It is noteworthy that
the early preon models did not, in general, provide
a more rational explanation of the quarks and leptons
of those days. Typically, four different preons were used to
explain the existence of the four particles of the
lightest quark/lepton family, while the heavier
ones were not properly understood. A good preon
model should hence explain {\it all} quarks and leptons
on equal terms, at least in some sense.

\section{Unstable Fundamental Particles?}
Most quarks and leptons are unstable, decaying into
lighter ones, until only stable (?) electrons, protons and
neutrinos remain. In my view there is a logical
problem with unstable fundamental particles:
How can Nature's most fundamental objects decay
into equally fundamental objects? As to the best
of my knowledge, this simple argument has not been
debated earlier in the literature. When discussing
this with colleagues, I am often met by the
attitude that Nature is anyway so strange,
so why bother? 

Again the History Book gives some moral support to the preon
solution, because all decays of previous ``fundamental'' particles
have mirrored their compositeness. Examples are:

\vskip0.1cm
- the decays of atoms were found to come from the decays of their nuclei;

- which were later blamed on the decays of their nucleons;

- which were later blamed on the decays of their quarks.
\vskip0.1cm

So, where will this sequence of explanations end? In my opinion,
not until we find a level with {\it stable} constituents. A preon model
must therefore either have absolutely stable preons, or rely
on yet unstable and composite preons (``pre-preons''). If preons
are stable, all quark and lepton decays are just
regrouping of preons into systems with lighter quarks and leptons.
This strongly limits the number of possible preon models.
The pioneering preon models quoted above did not have this
property. The heavier two families either remained unexplained,
or were assumed to be internal excitations of the lightest one,
decaying to their own ground states.

Summing up the prophecies so far, one can envisage the existence
of three absolutely stable preons. This makes {\it net preon flavour}
a completely conserved quantity in all particle reactions.
A by-product is that in case two quarks or leptons would have
identical net preon flavours, they will (or must) mix into
new mass eigenstates. This would be equivalent to, e.g.,
$\rho / \omega$- and $\eta / \eta'$-mixing in the quark model.
Hence such a preon model provides a possibility to understand
the various mixings of quarks, neutrinos and weak isospin
eigenstates in the SM. These aspects will be discussed later.

\section{Ad hoc Quantum Numbers}
The SM contains a few quantum numbers that have been put in
by hand, without a deeper understanding, and just in order
to describe that some quark and lepton quantities are conserved
or partially conserved.

These are quantum numbers that either seem absolutely conserved,
such as baryon number, or are known to be conserved only
in some reactions, e.g., weak isospin. Lepton numbers form a
grey-zone. They seemed to be absolutely conserved until
quite recently, when neutrino oscillations were discovered.

The History Book again gives some clues. A long time ago, the
``isotopic numbers'' were understood only after the discovery
of the neutron and the compositeness of atomic nuclei. In the 1950s
hadronic isospin and strangeness/hypercharge were introduced to
describe the observed approximate symmetries of hadronic decays and
interactions. These three quantum numbers turned out to come from
three quark flavours.

Since the number three pops up also in the SM, it is tempting
to guess that lepton number conservation has to do with preon
number conservation, i.e., preon-flavour $SU(3)$ symmetry.
The disturbing neutrino oscillations/mixings can then be
understood if two or more neutrinos have identical net
preon flavour. This will be discussed in the next Section.
The connection between weak isospin and the number of
preons is not as obvious, but will also be discussed below.

\section{Oscillations and Mixings of Fundamental Particles?}
The SM prescribes that certain created quarks, leptons and
gauge bosons mix into new eigenstates before being detected.
The History Book tells about a few such situations in the past.

A classical example is that different isotopes, created in
nuclear reactions, mix in Nature in certain proportions,
and are inseparable in normal chemical reactions. Chemical
isotope mixing hence has its root in the compositeness of atomic
nuclei.

A more modern case is the mixing/oscillation of the two neutral
kaons $K^{0}$ and $\bar{K}^{0}$ into the mass eigenstates
$K_{L}$ and $K_{S}$. This comes about because of quark
reactions inside the kaons, i.e., the compositeness of hadrons.

Both cases are examples of what happens when virtually
different states have identical net quantum numbers
relative the particular interaction used to ``detect''
them. I will now discuss three similar mixings
of ``fundamental'' particles in the SM, and their
interpretations in terms of compositeness.

\subsection{Example 1: The Cabibbo Mixing of $d$ and $s$}
Let us assume that the $d$ and $s$ quarks mix into the
weak-interaction (``mass'') eigenstates $d'$ and $s'$ 
because they have identical net preon contents. The
question is then how this can be arranged in detail.
There are in principle two different solutions. One is
that the two quarks have identical preon contents, although
with some internal differences between the detailed
preon wave functions, e.g., with two different internal spin
structures. The other one is to focus on the word
``net'', meaning that some preon flavours cancel inside
either quark, e.g., because they contain
preon-antipreon pairs.

It turns out that the first alternative fits the neutrino
sector better (see below). I therefore suggest that a
quark contains a preon-antipreon pair plus an additional
preon. As an example, consider a situation where the two
quarks have the following compositions:
$d = \beta \bar{\beta} \bar{\delta}$ and
$s = \alpha \bar{\alpha} \bar{\delta}$,
both with the net flavour of the $\bar{\delta}$ preon.
If we assume that quark production in strong (QCD)
processes always starts with pure preon states, these two
quarks will subsequently mix into mass
eigenstates before they decay, or take part in
a weak process. If looked upon as an
oscillation, the mixing can be illustrated with
the Feynman-like diagram of Fig. 2.
A preon-antipreon pair inside  a quark can annihilate
and produce another pair, thereby forming another quark.

\begin{figure}
\begin{center}
\includegraphics[width=.8\textwidth]{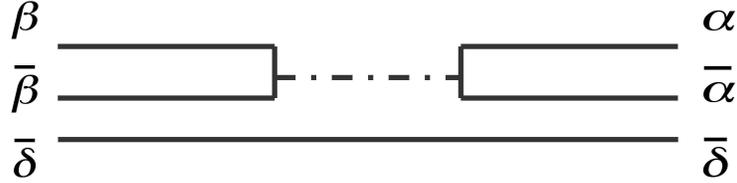}
\end{center}
\caption[]{One way for two composite quarks to mix
quantum-mechanically. A preon-antipreon pair annihilates
and turns into another pair}
\label{eps2}
\end{figure}

For technical reasons I have used the convention of an
antipreon ($\bar{\delta}$) for the common preon of $d$ and $s$.
It is naturally not clear in detail how one preon pair converts to
another. The intermediate system can be one or more photons,
one or more gluons, or maybe some new gauge bosons
(``hypergluons'').

Neither does it seem realistic that one and the
same phenomenon can explain {\it all} quark mixings
of the so-called CKM matrix \cite{cabibbo63,kobayashi73}.
There is simply no third state made up of these three
preons that can oscillate in a similar way to a $d$ or an $s$.
On the other hand, the smaller CKM matrix elements differ
by an order of magnitude from the Cabibbo one, indicating
that the detailed mechanism here is different
from the one of Fig. 2 for $d$ and $s$.
I refer to \cite{dugne02} for a preon-based discussion
of other quark mixings than the $d / s$ one.

The preon charges can easily be chosen as to
fit the charge of the two quarks. Obviously, $\delta$ must have
charge $+e/3$, while the other two can be ``anything''.
It is wise to choose the charges $+e/3$ also for
$\alpha$ and $-2e/3$ for $\beta$. This will be obvious from
the discussion in 6.2 below, although it already
now seems attractive to mimic the three charges of
the original flavour-$SU(3)$ quark model.

\subsection{Example 2: Neutrino Mixing}
Once we choose preon charges as multiples of $e/3$, and
prescribe that a quark contains three entities,
it becomes almost necessary to build the integer-charged
leptons as three-preon states. With the charges defined
above there are just three main ways to make neutrinos:
$\alpha \beta \delta$, $\alpha \alpha \beta$ and
$\delta \delta \beta$.

The three preons build up a total spin $1/2$,
which means that there are several possible spin combinations.
For simplicity, let us assume that two
unequal preons prefer to form a total spin-$0$ ``dipreon''
pair, in the same fashion as quarks tend to pair up
in diquarks in many situations \cite{anselmino93}.
Such restrictions result in the following {\it five}
neutrinos:
$\nu_{1} = \alpha(\beta \delta)_{S=0}$,
$\nu_{2} = \beta(\alpha \delta)_{S=0}$,
$\nu_{3} = \delta(\beta \alpha)_{S=0}$.
$\nu_{4} = \alpha(\beta \alpha)_{S=0}$,
$\nu_{5} = \delta(\beta \delta)_{S=0}$.
Note the similarity with the five neutral baryons in
the spin-$1/2$ nonet (octet + singlet) of the original quark model.

It can be seen that the three neutrinos $\nu_{1}$, $\nu_{2}$ and
$\nu_{3}$ indeed have identical net preon flavours, and hence
can mix/oscillate into three new mass eigenstates.
Such an oscillation can be seen as one of the three preons
oscillating between the two spin-$0$ pairs that it can form
with the other two preons. Figure 3 illustrates the situation
for an oscillation between $\nu_{1}$ and $\nu_{2}$. At this
primitive stage of the discussion it is not possible to
pinpoint the actual neutrinos that correspond to the different
preon states. Three of the five states must naturally
be connected to the three known neutrinos, while the
other two must have masses in excess of half the $Z^{0}$ mass.

\begin{figure}
\begin{center}
\includegraphics[width=.7\textwidth]{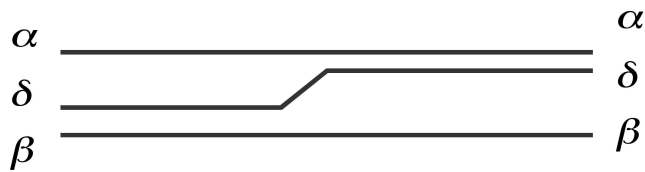}
\end{center}
\caption[]{One way for two composite neutrinos to
mix/oscillate. A preon oscillates in and out of two
different ``dipreon'' (spin-$0$) pairs}
\label{eps3}
\end{figure}

It can be added that with these particular preons it is impossible
to construct {\it charged} leptons that would oscillate or mix
in the same way. Neither can there be decays of the type
$\mu \rightarrow e \gamma$.

\subsection{Example 3: Electroweak Mixing of $W^{0}$ and $B^{0}$}
In many preon models \cite{souza92} the weak gauge bosons
are supposed to be composite and built up by preon-antipreon
pairs in total spin $1$. With the preons described earlier,
it is tempting to define $W^{+} = (\alpha \bar{\beta})$
and $W^{-} = (\beta \bar{\alpha})$.
The neutral sector is more complicated. There are five neutral
combinations of the three preons, and three of these have identical
net preon numbers: $\alpha \bar{\alpha}$, $\beta \bar{\beta}$ and
$\delta \bar{\delta}$. They resemble the $\rho^{0}$, $\omega$ and
$\phi$ of the vector meson nonet, and are expected to mix.
Looking closer into the preon-flavour $SU(3)$ structure, it seems
as if some (wave-function) combinations correspond to the
two neutral weak-isospin eigenstates, namely
$W^{0} = (\alpha \bar{\alpha} - \beta \bar{\beta})/\sqrt{2}$ and
$B^{0} = (\alpha \bar{\alpha} + \beta \bar{\beta})/\sqrt{2}$.
These two then mix into two mass eigenstates, $Z^{0}$ and
$Z'$ (a new and even heavier boson). Note that the mixing partner
of the normal $Z$ is {\it not} the photon, like in the SM.
The reason is that the weak interaction is not fundamental
in models with composite $Z$s and $W$s, and there can hence not
be an electroweak unification (and no Higgs!). This aspect will
be further discussed later.

The mixing of $(\alpha \bar{\alpha})$ with $(\beta \bar{\beta})$
can again can be illustrated in the fashion of a Feynman diagram,
as in Fig. 4.

\begin{figure}
\begin{center}
\includegraphics[width=.7\textwidth]{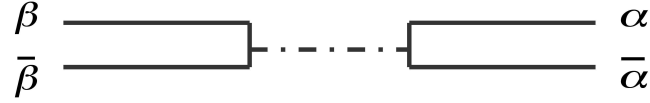}
\end{center}
\caption[]{One way for two composite neutral vector bosons
to mix. A preon-antipreon pair annihilates
and turns into another pair}
\label{eps4}
\end{figure}

The intermediate state must now be neutral in both charge and
colour, as well as have spin $1$. It could hence be one or more
photons, two or three gluons, or a number of hypothetical
``hypergluons''.

\subsection{An Interesting Comparison}
A closer look shows that the action in Figs. 2 \& 4 takes place
through very similar preon subprocesses, as shown again
in Fig. 5.

\begin{figure}
\begin{center}
\includegraphics[width=.7\textwidth]{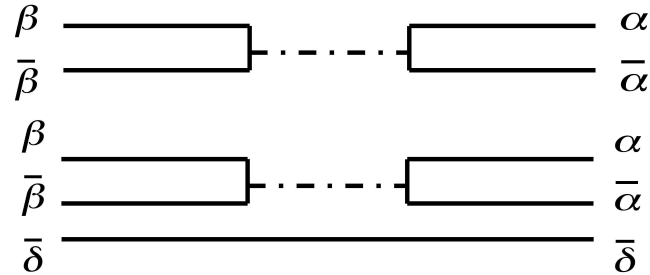}
\end{center}
\caption[]{The Weinberg mixing of $W^{0}$ and $B^{0}$, in the
upper part of the figure, and the Cabibbo mixing of $d$ and $s$,
in the lower part, come from the same basic preon processes,
and are hence related}
\label{eps5}
\end{figure}

If the mixing fraction of the
$(\alpha \bar{\alpha})$ and $(\beta \bar{\beta})$ pairs
depends {\it only} on the masses and/or electric charges of
$\alpha$ and $\beta$, and not on, e.g., the systems they sit in,
or their total colour or spin, then one can show that there
is a relation between the Cabibbo and Weinberg angles
\cite{dugne02}:

\begin{equation}
\cos\theta_{W} - \sin\theta_{W} = \sqrt{2}\sin\theta_{C}.
\end{equation}

With $\sin^{2}\theta_{W} = 0.23117 \pm 0.00016$ and
$\sin\theta_{C} = 0.2225 \pm 0.0035$
\cite{pdg02} the $lhs = 0.396 \pm 0.001$
and the $rhs = 0.315 \pm 0.005$. This is a
fair agreement considering the rough assumptions.
In addition, if we simplify the situation further and
assume that the mixings occur through one-photon
intermediate states, then the preon-antipreon pairs occur in
proportion to their squared charges, and we
expect:

\begin{equation}
\sin\theta_{C} = q^{2}_{\alpha}/(q^{2}_{\alpha}+q^{2}_{\beta}) =
1/5,
\end{equation}

\noindent which again is not far from reality.

\section{Massive, Charged and Unstable Weak Gauge Bosons}
Most {\it ad hoc} SM assumptions that lack a deeper background
have to do with the ugly fact that the weak interaction has massive
gauge bosons. That is the excuse for introducing the Higgs mechanism
with all its inherent problems and unsatisfactory logic. This
is typical for many of the odd features of the SM. They are
introduced to cure and/or quantify earlier theoretical problems,
rather than to explain the observationes from more basic
principles. Ofter the cure causes side-effects, which
need new cures, etc.

The situation with massive and unstable ``gauge bosons''
is not new in high-energy physics. In nuclear physics
it is often productive to regard the vector mesons,
$\rho$, $\omega$, $\phi$ etc, as gauge bosons of the
nuclear forces that keep nucleons in place.
They are good phenomenological couriers of QCD,
by leaking colour-neutral quark-antiquark
pairs between nucleons. It seems as if spin-$1$ particles
are better than scalar ones in faking true (massless)
gauge bosons.

The History Book therefore warns us that massive gauge
bosons are not fundamental, but some kind of ``neutral''
leakage of more basic forces. Hence $W$ and $Z$ might
well be false gauge bosons that just leak the basic
preon forces.

It would imply that there are {\it nine} different heavy vector
bosons (if we believe in preon-flavour $SU(3)$). Hence
six very heavy ones are yet to be discovered. Five of the
bosons are neutral ($Z$s) and four charged ($W$s). Three of
the $Z$s have identical net preon flavour and would mix,
since they are built up by $\alpha \bar{\alpha}$,
$\beta \bar{\beta}$ and $\delta \bar{\delta}$.
The remaining two $Z$s are mutual antiparticles,
being $\alpha \bar{\delta}$ and $\delta \bar{\alpha}$.

\section{Electroweak Unification and $\gamma / Z$ Mixing}
It remains to discuss why $Z^{0}$ seems to mix with the
photon, in a way that is well described by the electroweak sector
of the SM and parametrized by the Weinberg angle.

The History Book again tells why, because the situation is not new.
In the 1960s it was discovered that the photon sometimes
behaves like a hadron in interactions with nucleons.
It was suggested that its wave function has a hadronic component,
consisting mainly of the $\rho$ meson, but with some fraction
also of more massive spin-$1$ mesons at higher $Q^{2}$ values.
This idea was dubbed Vector Meson Dominance (VMD) \cite{schildknecht72}.
It was developed in great detail and used to understand, or at
least describe, a wealth of data. It is still a viable model
for the behaviour of virtual photons in medium-$Q^{2}$ reactions
with hadrons. Hence a photon is believed to couple directly
to a $\rho$, like in Fig. 6, when interacting with a hadron.

\begin{figure}
\begin{center}
\includegraphics[width=.7\textwidth]{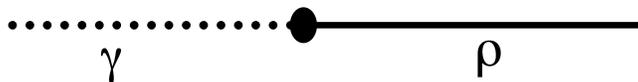}
\end{center}
\caption[]{The photon mixes with $\rho$ according to
the Vector Meson Dominance model}
\label{eps6}
\end{figure}

In the early days this was taken as a fact, and the origin
of the coupling remained unknown for a while. A physicist
with a gift to look into the future could then have come up
with a radical idea: The photon and the $\rho$ are one and
the same particle! This would lay the ground for a new
theory of {\it electrostrong unification}, where the
$\gamma / \rho$ mixing is parametrized by an electrostrong
mixing angle. After a while the substantial difference in
mass between the two particles would have inspired
someone to invent an electrostrong Higgs boson. Or rather,
a whole set of mixing angles and Higgs bosons, because
the photon couples also to other vector mesons.

Why was this model never invented (except here)? Because soon
after the birth of VMD, the quark model gained ground.
Then it became evident that any spin-$1$ neutral meson
couples to (``mixes'' with) the photon just because its
quarks are electrically charged. At high-enough $Q^{2}$
the photon sees the constituents, instead of the (invisible)
neutral hadron. The true explanation of the $\gamma / \rho$
mixing is hence {\it compositeness},
as illustrated by Fig. 7.

\begin{figure}
\begin{center}
\includegraphics[width=.7\textwidth]{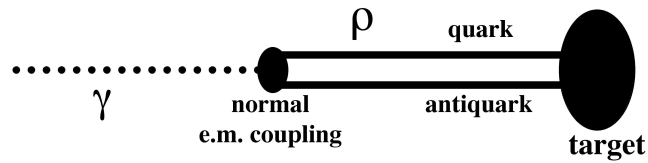}
\end{center}
\caption[]{The $\gamma / \rho$ mixing comes about because
$\rho$ is composite}
\label{eps7}
\end{figure}

The conclusion for the electroweak sector and preons is then obvious.
The $Z$ contains charged preons, and is hence destined to couple to a
high-$Q^{2}$ photon, as in Fig. 8.

\begin{figure}
\begin{center}
\includegraphics[width=.7\textwidth]{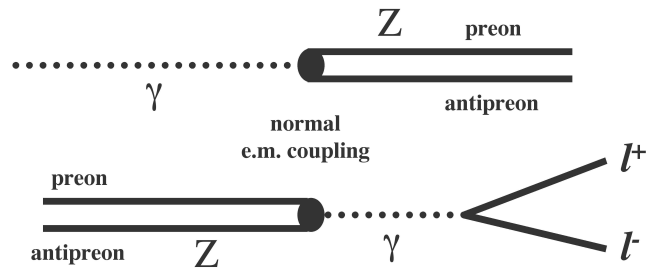}
\end{center}
\caption[]{The $\gamma / Z$ mixing comes about because
$Z$ is composite}
\label{eps8}
\end{figure}

Qualitatively, the electroweak formalism is very similar
to the old parametrization of the VDM model,
i.e., the propagators in a Feynman-graph formalism
look the same. The $Z$ and $W$s are just very good
at faking true gauge bosons. One can even understand why
one and the same Weinberg angle would reasonably
well describe not only the $\gamma / Z$ mixing in the SM
but also the $Z / Z'$ mixing in a preon model. This 
would come about if the conjectures leading to
(1) and (2) were true, i.e., that the preon-antipreon
states inside the various $Z$s mix via virtual photons,
rather than gluons or hypergluons.

However, the electroweak and electrostrong situation differ
quantitatively, since no heavier ($Z'$) boson
has revealed itself experimentally as
deviations from the SM predictions, e.g., at the
CERN LEP machine. Neither are there any traces of
direct $Z'$ production at the Fermilab Tevatron.
The hypothetical $Z'$ should therefore be much heavier
than the $Z^{0}$ \cite{pdg02}, unlike the vector-meson
situation, where the $\omega$ is very close to $\rho$
in mass. This need not be troublesome for preons,
because the preon-$SU(3)$ wave functions inside $Z$s
might differ substantially from the quark-$SU(3)$ wave
functions inside mesons.

It could be fruitful, however, to analyse the unconventional
alternative that the $Z^{0}$ and the next $Z'$ have almost
{\it equal} masses. Maybe the $Z^{0}$ wave function is dominated
by the $(\beta \bar{\beta})$ combination and the first $Z'$ by
$(\alpha \bar{\alpha})$. Then $e^{+}e^{-}$ annihilation would
produce mainly the $Z^{0}$, for two reasons: (i) an annihilation
to a photon would couple the photon four times as strongly to a
$\beta \bar{\beta}$ pair than to an $\alpha \bar{\alpha}$ pair;
(ii) an annihilation of only the dipreons in the $e^{+}e^{-}$
pair would result in a $\beta \bar{\beta}$ system (see Table 2
below), which in turn would prefer to make a $Z^{0}$. At present
it is unclear to me if a ``hidden'' $Z'$, near the $Z^{0}$ in mass,
would be in conflict with the CERN LEP precision tests of the SM.

\section{Conclusions: What next?}
This discussion of crucial {\it ad hoc} features of the
standard model has, step-by-step, led to strict and
rather detailed requirements of a hypothetical preon
model for quarks, leptons and heavy
vector bosons. As you might have guessed, they are all
in line with the ``preon-trinity'' model presented earlier
by Dugne, Hansson and myself \cite{dugne02}.

The three preons have the charges given in Table 1. Also listed
are the dipreons, necessary for understanding, e.g., neutrino
oscillations. When entered as anti-dipreons, the scheme reveals
a nice supersymmetric structure.

\begin{table}
\caption{A ``supersymmetric'' scheme of spin-$1/2$ preons
and spin-$0$ anti-dipreons.}
\begin{center}
\begin{tabular}{c|ccc}
charge
& ~$+e/3$
& $-2e/3$
& $+e/3$\\
\hline
spin-$1/2$ preons
& ~$\alpha$
& $\beta$
& $\delta$ \\
spin-$0$ (anti-)dipreons~~
& ~$(\bar{\beta} \bar{\delta})$
& $(\bar{\alpha} \bar{\delta})$
& $(\bar{\alpha} \bar{\beta})$\\
\end{tabular}
\end{center}
\end{table}

Following the SM prophecies, the leptons are now
constructed as combinations of one preon and one dipreon, the
quarks of one preon and one anti-dipreon, and the
heavy vector bosons of one preon and one antipreon,
as in Table 2.

\begin{table}
\caption{The composite states in the preon-trinity model;
leptons with one preon and one dipreon, quarks with one
preon and one anti-dipreon, and heavy vector bosons with
one preon and one antipreon.}

\begin{center}
\begin{tabular}{c|ccc|ccc|ccc|c}
& ~$(\beta \delta)$
& $(\alpha \delta)$
& $(\alpha \beta)$~
& ~$(\bar{\beta} \bar{\delta})$
& $(\bar{\alpha} \bar{\delta})$
& $(\bar{\alpha} \bar{\beta})$~
& ~$\bar{\alpha}$
& $\bar{\beta}$
& $\bar{\delta}$~  \\
\hline
$\alpha$~
& ~$\nu_{e}$
& $\mu^{+}$
& $\nu_{\tau}$~
& ~$u$
& $s$
& $c$~
& ~$Z^{0},Z'$
& $W^{+}$
& $Z^{*}$~
& ~$\alpha$ \\
$\beta$~
& ~$e^{-}$
& $\bar{\nu}_{\mu}$
& $\tau^{-}$~
& ~$d$
& $X$
& $b$~
& ~$W^{-}$
& $Z',Z^{0}$
& $W'^{-}$~
& ~$\beta$ \\
$\delta$~
& ~$\nu_{\kappa 1}$
& $\kappa^{+}$
& $\nu_{\kappa 2}$~
&~ $h$
& $g$
& $t$~
& ~$\bar{Z}^{*}$
& $W'^{+}$
& $Z'',Z'$~
& ~$\delta$   \\
\end{tabular}
\end{center}
\end{table}

The best way to test the model would be to search for the
new (heavy) particles in Table 2. Since the top
quark is most probably the one marked $t$,
there is some hope that the missing quarks and leptons
can be produced at least at the Fermilab Tevatron.
I am now analysing the conjecture that all three
quarks, $h$, $g$ and $t$, in Table 2 are hidden in the
Tevatron data. The analyses of their decay modes are rather
complex, especially in the light of the trigger and tagging
conditions of the Tevatron experiments.

The $X$ quark is a mystery. It appears near the $d$ and $b$
in Table 2, but it has charge $-4e/3$, since both its preon
and anti-dipreon have charge $-2e/3$. This made us suspect
\cite{dugne02} that the system is not bound, and hence not
a real quark.

A final observation is that the heavy neutrino
$\nu_{\kappa 2}$ in Table 2 can be produced in
$e^{+}e^{-}$ annihilation together with a {\it normal}
antineutrino. We expect the $\nu_{\kappa 2}$ to be
lighter than its related quark ($t$), since quarks are
in general heavier than the corresponding leptons.
Then the $\nu_{\kappa 2}$ can, in principle, be produced
at energies lower than around $175$ GeV. Hence it might
be visible in existing CERN LEP data. I am analysing the
situation in some detail. All LEP groups have reported 
that there are no statistically significant signatures
of decay products from a heavy neutrino
(``neutral excited lepton''). So maybe a careful
event-by-event check is the only hope of finding a
hint of a heavy neutrino at LEP (or HERA).

\vskip01.cm
\noindent {\bf Acknowledgements}\\ \\
First of all I would like to thank the Organisers for providing
a wonderful atmosphere at Schlo$\mbox{\greeksym{b}}$ Ringberg, and
a most fantastic weather all over Bavaria. Remember that this
praise comes right from the heart of a resident of the Swedish Arctic,
where the ice on the Baltic broke up in early May, six weeks ago!
I am also grateful for much advice from my preon partners
Jean-Jacques Dugne and Johan Hansson.
At an early stage this preon research was supported in part
by the European Commission under contract CHRX-CT94-0450,
within the network ``The Fundamental Structure of Matter".


\begin{thebibliography}{999}
\addcontentsline{toc}{section}{References}


\bibitem{dugne02}
J.-J.~Dugne, S.~Fredriksson, J.~Hansson:
Europhys. Lett. {\bf 57}, 188 (2002)

\bibitem{gellmann64}
M.~Gell-Mann: Phys. Lett. {\bf 8}, 214 (1964)

\bibitem{zweig64}
G.~Zweig: report CERN-TH-412 (1964), unpublished

\bibitem{harari79}
H.~Harari: Phys. Lett. {\bf 86B}, 83 (1979)

\bibitem{shupe79}
M.A. Shupe: Phys. Lett. {\bf 86B}, 87 (1979)

\bibitem{fritzsch81}
H.~Fritzsch, G.~Mandelbaum: Phys. Lett. {\bf 102B}, 319 (1981)

\bibitem{pdg02}
K.~Hagiwara et al.: {\it Review of Particle Physics},
Phys. Rev. {\bf 57}, 188 (2002)

\bibitem{souza92}
I.A.~D'Souza, C.S.~Kalman: {\it Preons} (World Scientific, Singapore 
1992)

\bibitem{cabibbo63}
N.~Cabibbo: Phys. Rev. Lett. {\bf 10}, 531 (1963)

\bibitem{kobayashi73}
M.~Kobayashi, T.~Maskawa: Prog. Theor. Phys. {\bf 49}, 652 (1973)

\bibitem{anselmino93}
For a review of diquarks, see M.~Anselmino, E.~Predazzi,
S.~Ekelin, S.~Fredriksson, D.B.~Lichtenberg:
Rev. Mod. Phys. {\bf 65}, 1199 (1993)

\bibitem{schildknecht72}
For a review of the early VMD model, see D.~Schildknecht:
Springer Tracts in Modern Physics {\bf 63}, 57 (1972)


\end{thebibliography}
\end{document}